\documentclass[12pt]{article}
\usepackage[dvips]{graphicx}
\usepackage{amssymb}
\usepackage{amsmath}
\usepackage{epsfig}
\usepackage{colortbl}

\numberwithin{equation}{section}
\numberwithin{table}{section}

\def\beq{\begin{equation}}
\def\eeq{\end{equation}}
\def\be{\begin{equation}}
\def\ee{\end{equation}}
\def\bea{\begin{eqnarray}}
\def\eea{\end{eqnarray}}

\def\jbar{\bar \jmath}
\def\ibar{\bar \imath}
\def\Re{{\rm Re\,}}
\def\Im{{\rm Im\,}}
\def\Mp{M_{\rm Pl}}
\def\P{{W_g}}

\DeclareRobustCommand{\SkipTocEntry}[4]{}
\RequirePackage{color}

\newcommand\diff{\mathrm{d}}
\def\vev#1{{\langle #1\rangle}}
\newcommand\e{\mathrm{e}}
\newcommand\iu{\operatorname{i}}

\begin{document}
\begin{titlepage}
\begin{center}
\rightline{\small ZMP-HH/13-16}
\rightline{\small CERN-PH-TH/2013-307}
\rightline{\small DESY-13-252}
%\rightline{draft: \today}
\vskip 2cm

{\Large  \bf
 On Moduli Spaces in  $\bf{AdS_4}$ Supergravity}
\vskip 1.2cm

{\large   Senarath de Alwis,$^{a}$ Jan Louis,$^{b,c}$ Liam
  McAllister,$^{d}$\\ \vskip 0.1cm Hagen Triendl,$^{e}$ and Alexander Westphal$^{f}$}

\vskip 0.8cm

$^{a}${\em UCB 390, Physics Department, University of Colorado, Boulder CO 80309, USA}
\vskip 0.2cm

$^{b}${\em Fachbereich Physik der Universit\"at Hamburg,\\ Luruper Chaussee 149, 22761 Hamburg, Germany}
\vskip 0.2cm

{}$^{c}${\em Zentrum f\"ur Mathematische Physik,
Universit\"at Hamburg,\\
Bundesstrasse 55, D-20146 Hamburg, Germany}
\vskip 0.2cm

$^d${\em Department of Physics, Cornell University, Ithaca, NY 14853, USA}

\vskip 0.2cm

$^e${\em Theory Division, Physics Department, CERN, CH-1211 Geneva 23, Switzerland}

\vskip 0.2cm

$^f${\em Deutsches Elektronen-Synchrotron DESY, Theory Group, D-22603 Hamburg, Germany}

\vskip 0.3cm

{\tt dealwiss@colorado.edu, jan.louis@desy.de, mcallister@cornell.edu, hagen.triendl@cern.ch, alexander.westphal@desy.de}

\end{center}

\vskip 1cm

\begin{center} {\bf ABSTRACT } \end{center}

\noindent

We study the structure of the supersymmetric moduli spaces of ${\cal N}=1$ and  ${\cal N}=2$
supergravity theories in $AdS_4$ backgrounds.
In the ${\cal N}=1$ case, the moduli space cannot be
a complex submanifold of the K\"ahler field space, but is instead
real with respect to the inherited complex structure.
In ${\cal N}=2$ supergravity the same result holds for the vector multiplet moduli space, while the hypermultiplet moduli space is a K\"ahler submanifold of the quaternionic-K\"ahler field space.
These findings are in agreement with  AdS/CFT considerations.

\vfill

\bigskip

%December 2013
%\today

\end{titlepage}

%%%%%%%%%%%%%%%%%%%%%%%%%%%%%%%%%%%%%%%%%%%%%%%%%%%%

%\tableofcontents

%%%%%%%%%%%%%%%%%%%%%%%%%%%%%%%%%%%%%%%%%%%%%%%%%%%%

\section{Introduction}

Vacua of supersymmetric field theories  and supergravities frequently have
continuous degeneracies parameterized by the
background values of one or more scalar fields.
The structure and properties  of these moduli spaces depend on the amount of supersymmetry, on the spacetime background, and on whether supersymmetry is a global or local symmetry.

The existence of
moduli spaces in  supersymmetric compactifications of string  theory to four dimensions impedes the construction of realistic models of particle physics and cosmology, and a primary endeavor in string phenomenology is the study of mechanisms that lift the vacuum degeneracy and stabilize the moduli.
Moduli spaces of $AdS_4$ vacua are rather different from the  better-understood moduli spaces of supersymmetric Minkowski solutions, and it is worthwhile to discuss their special properties.

In this paper we will focus on the structure of  supersymmetric moduli spaces in $AdS_4$ vacua of general ${\cal N}=1$ and ${\cal N}=2$ supergravities.\footnote{For  early work on $AdS$  supersymmetry, see \cite{Keck:1974se,Zumino:1977av} and the lectures \cite{deWit:1999ui}. Recent results in  global ${\cal N}=1$  supersymmetry in $AdS_4$ include \cite{Adams:2011vw,Festuccia:2011ws,Aharony:2012jf}.}
Two  elementary  structural questions are whether a continuous moduli  space exists, and  when one does,  how the moduli space geometry is related to that  of the parent configuration space.
To frame the question, we  compare to the
simpler case of a supersymmetric Minkowski background $M_4$.
Consider an ${\cal N}=1$ theory with global or local  supersymmetry, containing $n_c$  complex scalar fields $\phi_i$, $i=1,\ldots, n_c$.
The scalars parameterize a K\"ahler manifold ${\cal C}$.
The supersymmetric vacua of this theory in a Minkowski background $M_4$ are determined by the solutions of $n_c$ holomorphic
equations $\partial_i W=0$, where the superpotential $W$ is a holomorphic
function of the complex scalars $\phi_i$.
Generically the solutions are isolated points, but when a continuous moduli space ${\cal M}$ exists, it
 is a complex, and therefore K\"ahler,  submanifold of
${\cal C}$.

The situation is quite different in $AdS_4$.  For an $AdS_4$ vacuum of ${\cal N}=1$  supergravity,  we will see that ${\cal M}$ cannot be
a complex submanifold of ${\cal C}$: instead, ${\cal M}$ is real with respect to the inherited complex structure,
and can at best have real dimension $n_c$, i.e.~half the dimension of  the parent configuration space ${\cal C}$.
In an $AdS_4$ vacuum of ${\cal N}=2$ supergravity, ${\cal C}$ is the product of a
special K\"ahler  manifold and a quaternionic K\"ahler  manifold, and we will show that the moduli space ${\cal M}$
is again a submanifold of ${\cal C}$, consisting of a real manifold times a K\"ahler manifold  -- also of  at most
half the parent dimension.

An intuition for the structure of the moduli spaces
comes from the AdS/CFT correspondence,
which relates an $AdS_4$ background to a three-dimensional superconformal
field theory (SCFT) on the boundary.
For ${\cal N}=1$ in the bulk one has four supercharges, leading to a superconformal
$N=\tfrac12$ theory  on the  three-dimensional boundary.\footnote{By
  $N=\tfrac12$ we mean a three-dimensional theory with two ordinary
  supercharges, or four supercharges in the superconformal case.} In
this case  each chiral multiplet has
only one real scalar,
and thus one  can find at most a
moduli space of half the original dimension.
For ${\cal N}=2$ in the bulk one has an ${\cal N}=1$ theory on the three-dimensional
boundary which can feature chiral and vector multiplets.
In this case ${\cal M}$ is a real manifold with a K\"ahler submanifold.
Upon dualizing the three-dimensional vectors,
 ${\cal M}$ becomes K\"ahler.

This paper is organized as follows:
in section \ref{sec:N1} we  determine  the structure of the moduli space in theories
with ${\cal N}=1$ supersymmetry, while section \ref{sec:N2} extends our analysis to ${\cal N}=2$.
In Appendix \ref{globalAdS}  we discuss the global limit of  ${\cal N}=1$ supergravity in $AdS_4$, and  Appendix \ref{moreexamples} contains a few illustrative examples.
In the main text we set  the reduced  Planck mass $M_{\rm{pl}}$  to unity, but we retain explicit factors of $M_{\rm{pl}}$ in the discussion of decoupling in Appendix  \ref{globalAdS}.

\section{${\cal N}=1$ Supergravity in $AdS_4$}\label{sec:N1}

In ${\cal N}=1$ supergravity the scalar fields $\phi^i,\, i=1,\ldots,n_c$
are members of chiral multiplets with the
two-derivative
Lagrangian \cite{Wess:1992cp}
\begin{eqnarray}\label{N=1Lagrangian}
{\cal L}\ =\ -\tfrac12 R - K_{i\jbar} \partial_\mu \phi^{i} \partial^{\mu} \bar\phi^{\jbar}
- V(\phi,\bar\phi) \ ,
\end{eqnarray}
where $R$ is the scalar curvature,
$K_{i\jbar} = \partial_i \bar\partial_{\jbar} K$ is a K\"ahler metric on the scalar field space ${\cal C}$,
with K\"ahler potential $K$, and the scalar potential $V$ is given
by
\begin{equation}\label{eq:V1}
V=e^{K}(K^{i\jbar}D_{i}W D_{\jbar}\bar{W}-3|W|^{2})\ ,\qquad
\textrm{with}\qquad D_{i}W= \partial_i W + K_i W\ .
\end{equation}
Supersymmetric minima  occur where
\begin{equation}\label{eq:DW=0}
D_{i}W  =0 = \bar D_{\ibar}\bar W\ \quad \forall~ i\ ,
\end{equation}
and the moduli space ${\cal M}$ is defined as the locus in ${\cal C}$ on which
\eqref{eq:DW=0} holds.\footnote{Here we neglect the possibility of having charged scalars and associated D-terms: these do not affect our analysis of the structure of the moduli space, as we will see shortly.}  We will use $\langle~~\rangle$  to denote  evaluation on  ${\cal M}$, so that
$\langle D_{i}W\rangle = \langle\bar D_{\ibar}\bar W\rangle\ = 0$ by definition, for all $i$.

\subsection{Structure of the moduli space}\label{N1modspace}

From \eqref{eq:V1} one infers that for
$\langle D_{i}W\rangle=\langle W\rangle=0$
the background is Minkowski space $M_4$, while for
$\langle W\rangle\neq0, \langle D_{i}W\rangle=0$ it is
$AdS_4$. Therefore the supersymmetric minima
in $M_4$ are determined by the %$n_c$
holomorphic equations
$\partial_i W=0$,
which are independent of
the K\"ahler potential $K$. For generic $W$,
these $n_c$ equations determine the $n_c$ complex variables $\phi^i$, leaving  no  continuous moduli space:  the vacuum  manifold is a set of isolated points.
On the other hand, for non-generic superpotentials ($W=0$ being a simple example), there can be a  continuous moduli space ${\cal M}$.
Because ${\cal M}$ is determined by the solution to  a set of holomorphic equations, it is a complex submanifold of the (K\"ahler) field space ${\cal C}$, and so ${\cal M}$  is K\"ahler.

The situation is different in $AdS_4$, because for $\langle W\rangle \neq 0$  the F-flatness conditions $D_{i}W=0$ depend on the K\"ahler potential, which is non-holomorphic.\footnote{Even in {\it{global}} $AdS_4$ supersymmetry, the superpotential transforms under K\"ahler transformations, as explained in \cite{Adams:2011vw,Festuccia:2011ws}.}
The equation counting  is unchanged, so it is still true that for generic $K$ and $W$,  the vacuum  solutions are isolated points.
However,  when a moduli space does arise  due to non-generic $K$ and $W$, its properties are different from the moduli spaces in  Minkowski solutions, as we now show.

In order to find the moduli space we infinitesimally vary the equations
\eqref{eq:DW=0} to obtain
\beq\begin{aligned}\label{DW=0sol}
&\langle \partial_j D_i W\rangle\, \delta\phi^j + \langle K_{i\jbar} W\rangle\,\delta\bar\phi^{\jbar}=0\ ,\\
&\langle K_{\ibar j}\bar W\rangle\,\delta\phi^j + \langle \bar \partial_{\jbar}\bar D_{\ibar}W\rangle\,
\delta\bar\phi^{\jbar}=0 \, .
\end{aligned}
\eeq
In matrix form we then have
\beq\label{Mform}
\mathbb{M}\
\left(\begin{array}{c}
\delta\phi^j\\
\delta\bar\phi^{\jbar}
\end{array}\right) = 0\ , \qquad \textrm{with}\qquad
\mathbb{M} =
\left(\begin{array}{cc}
m_{ij}& \langle K_{i\jbar}W\rangle\\
\langle K_{\ibar j}\bar W \rangle& \bar m_{\ibar \jbar}
\end{array}\right)\ ,
\eeq
where $ m_{ij} $
is proportional to the mass matrix of the fermions in the chiral multiplets, and is
given by \cite{Wess:1992cp}\footnote{The proportionality factor is $e^{K/2}$,
and  it is the matrix $e^{K/2} m_{ij}$ that appears in the Dirac equation. We also have made use of the fact that on ${\cal M}$, partial derivatives and covariant derivatives are equivalent, cf.~\eqref{eq:DW=0}.}
\beq\label{mre}
m_{ij} = \langle \nabla_iD_j W
\rangle
= \langle \partial_iD_j W
\rangle
= \langle \partial_i\partial_j W + K_j \partial_i W + K_{ij}W
\rangle
=\langle \partial_i\partial_j W +  (K_{ij}- K_iK_j)\, W
\rangle\ .
\eeq
Note that in an $M_4$ background $m_{ij}$ is just the second derivative of the superpotential, while in an $AdS_4$ background $\langle W\rangle \neq 0$ and thus the cosmological constant contributes.
Since  $\langle K_{i\jbar}\rangle$ is necessarily a positive matrix, and $\langle W\rangle\neq0$
in $AdS_4$,  we  learn that
the matrix $\mathbb{M}$ in \eqref{Mform}
has at least real rank $n_c$, leaving at most a moduli space
of real dimension $n_c$.\footnote{This follows from the fact that the matrix $\langle K_{i\jbar}W\rangle$ is invertible in an AdS vacuum: it therefore contains $n_c$ linearly independent vectors. Thus $\mathbb{M}$ also contains at least $n_c$ linearly independent vectors, so the rank of $\mathbb{M}$ is at least $n_c$.}

Before we proceed, let us note that including a D-term does not change the analysis. Gauging isometries on a K\"ahler manifold results in a D-term
of the form $D= k^i K_i$ where $k^i$ is an appropriately normalized Killing vector and  $K_i$ is the first derivative of the K\"ahler potential.
Gauge invariance of the superpotential further imposes $k^i \partial_i W=0$,
which implies that in $AdS_4$ the D-term
can alternatively be written as \cite{Gates:1983nr}
\begin{equation}
D=W^{-1} k^i D_i W \,.
\end{equation}
This expression shows that on ${\cal M}$ the D-term vanishes automatically,
and its variation,
\begin{equation}
\delta D = W^{-1} k^i \delta(D_i W) \,,
\end{equation}
is proportional to the variation of $\delta(D_i W)$ which  we already
analyzed in eq.~\eqref{DW=0sol}. Thus, the D-term imposes no further constraints on the moduli space.

To examine the structure of the moduli space, we rewrite \eqref{Mform} in terms of  real variations
obtained from the decomposition $\phi^i = \frac1{\sqrt{2}}(A^i+\iu B^i)$.
In this case, after choosing ${\rm{Im}}(\langle W \rangle)=0$, we
have
\begin{equation}
\mathbb{M}_{r}\left(\begin{array}{c}
\delta A^{j}\\
\delta B^{j}
\end{array}\right)=0\ ,\qquad\mathbb{M}_{r}=\left(\begin{array}{cc}
\langle\Re m_{ij}+K_{i\jbar}W\rangle & -\langle\Im m_{ij}\rangle\\
\langle\Im m_{ij}\rangle & \langle\Re m_{ij}-K_{i\jbar}W\rangle
\end{array}\right)\ .\label{eq:Mform2}
\end{equation}
We now observe that the complex structure on the space
of chiral fields in the given basis is
\begin{equation}
J=\begin{pmatrix}0 & -I\\
I & 0
\end{pmatrix}\quad,\quad J^{2}=-\begin{pmatrix}I & 0\\
0 & I
\end{pmatrix}\label{eq:J}
\end{equation}
where $I$ is the $ $$n_{c}\times n_{c}$ unit matrix. For the non-trivial
solution space of \eqref{eq:Mform2} (i.e.~the kernel of the map $\mathbb{M}_{r}$) to have the complex structure inherited from ${\cal M}$,
%requires that
the existence of a non-trivial solution to
\begin{equation}
\mathbb{M}_{r}\left(\begin{array}{c}
\delta A^{i}\\
\delta B^{i}
\end{array}\right)=0\label{eq:realsol}
\end{equation}
must imply that there is a non-trivial solution to
\[
\mathbb{M}_{r}J\left(\begin{array}{c}
\delta A^{i}\\
\delta B^{i}
\end{array}\right)=0\quad .
\]
But this means that
\begin{equation}
(J\mathbb{M}_{r}-\mathbb{M}_{r}J)\left(\begin{array}{c}
\delta A^{i}\\
\delta B^{i}
\end{array}\right)=0\quad .\label{eq:commutator}
\end{equation}
However, since
\[
J\mathbb{M}_{r}-\mathbb{M}_{r}J=2\left(\begin{array}{cc}
0 & \langle K_{i\jbar}W\rangle\\
\langle K_{i\jbar}W\rangle & 0
\end{array}\right)\quad ,
\]
and $\langle K_{i\jbar}W\rangle$ is non-singular,  it follows from \eqref{eq:commutator}
that only the trivial solution exists.  Thus, no nontrivial
solution space of \eqref{eq:realsol} can be complex in the
complex structure \eqref{eq:J}.

This result can be seen more explicitly.
Suppose there is a complex flat direction, say along the $1$ direction,
and consider fluctuations along this direction,  setting $\delta A^{i}=\delta B^{i}=0$ for $i \neq 1$.
From eq.~\eqref{eq:Mform2} we obtain
\begin{eqnarray}
\langle\Re m_{i1}+K_{i\bar{1}}W\rangle\delta A^{1}-\langle\Im m_{i1}\rangle\delta B^{1} & = & 0\quad,\label{eq:realeqn1}\\
\langle\Im m_{i1}\rangle\delta A^{1}+\langle\Re m_{i1}-K_{i\bar{1}}W\rangle\delta B^{1} & = & 0\quad,\label{eq:realeqn2}
\end{eqnarray}
which should hold for all $i$. Since $K_{i\bar \jmath}$ has rank $n_c$, there is at least one index $j_{\star}$  for which
$K_{j_{\star}\bar{1}}\neq 0$.
%\lm{fixed above} $K_{j_{\star}1}\neq 0$.
Taking $i=j_{\star}$, eq.~\eqref{eq:realeqn1} holds for arbitrary $\delta A^1 , \delta B^1$ only if $\Im m_{j_{\star}1}=0$ and $K_{j_{\star}\bar 1}W=-\Re m_{j_{\star}1}$.
Then eq.~\eqref{eq:realeqn2} would require that $\delta B^1=0$, negating the existence of a complex moduli space.

We conclude that the two scalars $(A,B)$ of a chiral multiplet
cannot simultaneously be  massless moduli. In other words,
the moduli space is necessarily real with respect to the original complex structure of the chiral multiplets.\footnote{Of course it is possible that an even number of
real moduli can be combined into complex fields with respect to another complex structure.}

It is instructive to compute the mass matrix of the scalar fields.
The first derivative of $V$ reads\footnote{We define the K\"ahler covariant derivative acting on a tensor to be $D_i=\nabla_i+K_i$ .}
\begin{equation}
\partial_k V= \nabla_k V = e^{K}\Big( K^{i{\jbar}}D_kD_iW{\bar D}_{\jbar}{\bar W}-2(D_kW){\bar W}\Big)\ ,
\label{eq:Vp}
 \end{equation}
which indeed vanishes at the minimum, where $\langle D_{i}W\rangle=0$.
From \eqref{eq:Vp} we can compute the bosonic mass matrix
\beq\begin{aligned}
\langle \nabla_k \nabla_{\bar l} V\rangle &= e^K \big(K^{i\jbar}
m_{ki}\bar{m}_{\bar l\jbar}
-2 K_{k\bar l}  |W|^2 \big)\ ,\\
\langle \nabla_k \nabla_{l} V\rangle &=-e^K m_{kl}\bar W\ ,
\end{aligned}
\eeq
where $m_{ki}$ is defined in \eqref{mre}.
Decomposing $\phi^i = \frac1{\sqrt 2}(A^i+\iu B^i)$ one obtains the
mass matrices for $A^i$ and $B^i$,
\beq\begin{aligned}\label{ABmass}
(m_{A}^{2})_{kl} & = e^K \big(K^{i\jbar} (m_{ki} -\tfrac12 K_{\bar k i} W )
(\bar{m}_{\bar l\jbar}- \tfrac12 K_{l\jbar} \bar W)
-\tfrac94 K_{k\bar l}  |W|^2\big)\ ,\\
(m_{B}^{2})_{kl} & =  e^K \big(K^{i\jbar} (m_{ki}+\tfrac12 K_{\bar k i} W )
(\bar{m}_{\bar l\jbar}+ \tfrac12  K_{l\jbar}\bar W)-\tfrac94 K_{k\bar l}  |W|^2\big)\ ,\\
(m^2_{AB})_{kl} & = 2 e^K\Im (m_{kl}{\bar W})\ .
\end{aligned}\eeq
On diagonalizing these equations one finds that
only one of the two real scalars in a chiral multiplet can be massless. We relegate the details of this discussion, as well as the relation to the rigid AdS limit and to the formulae of \cite{deWit:1999ui},
to Appendix \ref{globalAdS}.

\subsection{Examples of moduli spaces in $AdS_4$ supergravity}\label{Examples}

To illustrate the general results above, we consider a few examples.
Take $n_c=1$ and
\beq
K=\tfrac{1}{2}(\phi+\bar{\phi})^{2}\ ,\qquad  W=c = \textrm{constant.}\
\eeq
The F-term is $D_{\phi}W=(\phi+\bar{\phi})c$, which vanishes for $\Re\phi=0$.
We see that ${\cal M}$ is the locus  $\Re\phi=0$, on which the scalar potential is $\langle V\rangle=-3|c|^{2}$.
Thus, $\Im\phi$ is  an (axionic) flat direction parameterizing the moduli space.

As a slightly more involved example motivated from string theory, consider
 $p$ chiral fields $T$ and $q$ chiral fields $Q$ (i.e.\ $n_c=p+q$)
with couplings
\beq\begin{aligned}
K & =  K(T,\bar{T}) + Z(T,\bar{T})Q\bar{Q}+ {\cal O}(Q^3)\ ,\\
W & =  c+m(T) Q^{2}+{\cal O}(Q^{3})\ ,
\end{aligned}\eeq
where  $K(T,\bar{T})$ and $Z(T,\bar{T})$ are for the moment arbitrary real
functions of $T$ while $m(T)$ is an arbitrary holomorphic function.
The supersymmetry condition for $Q$ reads
\beq
D_{Q}W=  2mQ+Z\bar{Q}W+{\cal O}(Q^{2})
\eeq
which is solved by $Q=0$.  On the locus where $Q=0$, we have $\partial_{T}W|_{Q=0}=0$, so that ${\cal M}$ is the space of solutions of
\beq \label{Tdirection}
D_{T}W|_{Q=0}=K_{T}\,c=0\,.
\eeq
Because the condition (\ref{Tdirection})  depends only on $K(T,\bar T)$,  the functions $Z(T,\bar{T})$ and $m(T)$  are unconstrained.
For generic $K$, all $T$ are fixed by \eqref{Tdirection},
leaving no moduli space.
However,  moduli spaces arise in special cases: e.g.~for $K= K(T+\bar T)$ only the $\Re T$ are fixed, leaving the $p$ $\Im T$  directions
as axionic moduli.
As anticipated, the moduli space is real, with dimension at most half the dimension of the original K\"ahler manifold.
The background value of the potential is again $\langle V\rangle =-3|c|^{2}$.
Note that not every function $K= K(T+\bar T)$ is  compatible with the existence of a moduli space: an additional requirement is that $K$ is non-singular at $K_{T}=0$.
For example, the K\"ahler  potential $K=-\ln(T+\bar T)$, which is commonplace in tree-level  effective actions of string compactifications,
has $K_{T}=0$ only at $(T+\bar T)\to\infty$,
corresponding to an infinite and thus unacceptable $K$.
On the other hand, a  general polynomial $K= \sum_{n=2}^\infty a_n (T+\bar T)^n$  yields a moduli space.

We will  next discuss a simple example with a Goldstone-type global
$U(1)$ symmetry of the full Lagrangian. Let us choose
 $K$ and $W$ to be of the form
\begin{equation}\begin{aligned}
K =  \phi_{1}\bar{\phi}_{1}+\phi_{2}\bar{\phi}_{2}\ ,\quad & W=c+m \phi_{1}\phi_{2}\ ,\\
\end{aligned}\end{equation}
with $m$ and $c$ being real for simplicity.
In Appendix \ref{moreexamples} we show that there are the following two
supersymmetric solutions of  $D_{\phi_1}W= D_{\phi_2}W=0$:
\begin{itemize}
\item[A)] $\vev{\phi_{1}}=\vev{\phi_{2}}=0$ and
\item[B)] for $|c| >|m|$,  non-trivial solutions
with $\vev{\phi_1}=\pm \vev{\bar\phi_2}\neq0$.
\end{itemize}
In both cases the
symmetry $\phi_{1}\rightarrow e^{i\theta}\phi_{1},\,\phi_{2}\rightarrow e^{-i\theta}\phi_{2}$ is unbroken.
However, in the first solution no flat direction exists, while if we parameterize
$\phi_{1}= r_1 e^{i(\chi+\rho)},\,\phi_{2}= r_2 e^{i(\chi-\rho)}$, we see that $\rho$ is a flat direction in the second solution.

\subsection{Global symmetries and exact moduli spaces}\label{quantum}

In the examples just discussed, translation along the moduli space corresponds to a continuous shift symmetry.
%display the obvious fact that
%the moduli space always has a continuous
%translational isometry (shift symmetry) of the form
%\beq\label{shift}
%\delta T = i\gamma\ ,\qquad \gamma\in\mathbb{R}\ .
%\eeq
%However, this does not have to be a symmetry of the full theory
%but only needs to appear on the critical locus,
%as is the case in the second example.
%This feature might not
%be immediately  obvious from the form of $W$, and can therefore be difficult
%to check or control.
%To summarize, a sufficient (but not necessary) condition for an AdS moduli space
%is that $K$ and $W$ have a shift symmetry of the form \eqref{shift}.
%A necessary (but not sufficient) condition is that $K$
%has the shift symmetry \eqref{shift}.
However, well-known arguments exclude exact continuous global symmetries in string theory, and in quantum gravity  more generally (see e.g.~\cite{Banks:1988yz,Abbott:1989jw,Kallosh:1995hi}, and the recent discussion in \cite{Banks:2010zn}),  and one might ask whether these no-go results constrain the existence of exact quantum moduli spaces in quantum gravity theories.
To explain why there  is no associated constraint,  we begin by  briefly recalling  two of the  standard arguments.

\vskip 4pt
\noindent
{\it Black holes and global symmetries.}---Consider a  continuous global internal symmetry ${\cal G}$  under which  one or more species of particles, all with nonzero mass, are charged.  Denote by $\lambda_{max}$ the maximum ratio of ${\cal G}$-charge $q$ to mass $m$,  across all species in the spectrum  (not only the lightest species).  Form a  macroscopic Schwarzschild black hole  from constituents of total ${\cal G}$-charge $Q$ and mass $M_0$.  Then  once Hawking radiation  causes the black hole to decay to mass $M< M_{\star} \equiv Q/\lambda_{max}$,  it is not possible  for any subsequent decay process to release a total charge $Q$  while remaining consistent with conservation of energy.   So ${\cal G}$-charge is not conserved.  Note that one can make the initial black hole as large as  necessary in order to ensure that the Hawking temperature  remains  as small as desired when the black hole has mass $M_{\star}$,  so that semiclassical  reasoning remains valid.

The possibility of this process implies that in an effective theory derived from a consistent quantum gravity theory with standard black hole thermodynamics,  there must be operators violating every\footnote{Violation of axionic symmetries by black holes  (and by wormholes \cite{Abbott:1989jw}) is somewhat subtle, in part because of the possibility of axionic hair: see e.g.~\cite{Bowick:1988xh,Bowick:1989xg}.} continuous global internal symmetry.

%These operators may involve no derivatives, or they may involve couplings to spacetime curvature,  as we discuss below.
%Semiclassical black arguments establish that global symmetries are violated,
%but they neither exclude the possibility that this violation occurs only through higher derivatives,  nor do they  necessitate that this violation occurs only through higher derivatives.  The semiclassical argument is agnostic.

\vskip 4pt
\noindent
{\it String theory and global symmetries.}---Banks and Dixon showed in \cite{Banks:1988yz}  that for any exactly conserved  non-axionic global internal symmetry, one can construct a vertex operator for a gauge boson from the conserved global symmetry current.
This implies that  any exact  non-axionic global  internal\footnote{The  qualifier `internal'  is necessary because  Lorentz symmetry of noncompact spacetime is an exception to the argument of \cite{Banks:1988yz}, but this will not be relevant for our discussion.} symmetry must be gauged in string theory.  Axionic shift symmetries of the form
\begin{equation}
a \mapsto a+ const.  \label{axionshift}
\end{equation}
are not constrained by this argument.   At zero momentum, the vertex operator  for an axion $a$  is a worldsheet total derivative, and the worldsheet fields  do not transform under (\ref{axionshift}).
Thus, the logic of \cite{Banks:1988yz} does not apply to axionic symmetries, including translations along  the flat directions in the  first two examples above.

%The fact that axionic symmetries evade the above arguments  makes it clear that exact moduli spaces  protected by axionic symmetries are compatible with known results about symmetries in quantum gravity.   However, axionic symmetries are not the only possibility.
\vskip 4pt
\noindent
{\it Accidental symmetries.}---In view of the above  arguments, moduli spaces protected by exact, non-axionic global symmetries  are incompatible with general  reasoning about quantum gravity and string theory.   However, it is crucial to recognize that the presence of an exact moduli space {\it{does not}} imply the existence of any exact symmetry of the full Lagrangian:  it is consistent for the symmetry of translations along the moduli space to be an accidental symmetry  that is preserved along some locus.
Such a symmetry does not correspond  to a  current that is conserved at all points in the configuration space,  and so is not constrained by either of the arguments above.

Accidental symmetries  that hold only on a special locus in the  configuration space  can be broken by  non-derivative couplings or by derivative couplings.  To give two examples, the symmetry of translations $\phi \mapsto \phi+const.$   could be broken by
\begin{equation}
\Delta {\cal L} = \phi^2\chi^2\ ,
\end{equation} where $\chi$  is another scalar field,  or by
\begin{equation}
\Delta {\cal L} = \phi^2 R^2\ , \label{curvaturecoupling2}
\end{equation} where $R$ is the scalar curvature of spacetime.
The latter case is particularly relevant: there  is no conserved current in the full theory, but in a  Minkowski solution $\phi$  enjoys the accidental shift symmetry $\phi \mapsto \phi+const$, and hence an exact moduli space,  while in AdS  solutions  the coupling (\ref{curvaturecoupling2}) gives $\phi$  a mass  and lifts the  corresponding moduli space.

A particular  form of symmetry breaking generalizing (\ref{curvaturecoupling2}) arises in certain extended supergravities and in string theory compactifications with extended supersymmetry: the quantum gravity effects that destroy global charges and  hence prevent the associated global symmetries from being exact only appear beyond the level of the two-derivative action.
Examples include  compactifications with ${\cal N}=4$ and ${\cal{N}}=8$ supersymmetry, such as compactifications on $T^6$ of heterotic  string theory and type II  string theory, respectively.  In these compactifications
there is an exact continuous $SO(6,22)$ or $E_{7(7)}$
%\lm{are these two global symmetries ordered properly, corresponding to heterotic on $T^6$ and type II on $T^6$, respectively?} \Hnote{I corrected it.}
global symmetry group  at the level of ungauged supergravity, but the continuous symmetries are broken by instantons (e.g.\ D-instantons), i.e.\ by the charge lattice of the theory. These instantons contribute only at four and more derivatives, due to supersymmetry, and break the continuous symmetry group to a lattice corresponding to the monodromies of the charges of the theory.
In  theories of this sort where higher-derivative couplings  are the only effect  spoiling a symmetry, translation along an exact quantum moduli space  can then correspond to a  genuine exact symmetry of the two-derivative theory, which is only an accidental symmetry of the full theory incorporating higher derivatives.

To summarize,  we inquired whether the presence of an exact moduli space implies the existence of a symmetry that is forbidden by quantum gravity arguments.  It does not: the symmetry of translations along the  moduli space might be an  accidental symmetry of the full theory, preserved on some  special locus in the configuration space, hence avoiding no-go results from quantum gravity.
%The operators  breaking the symmetry away from the special locus  may involve higher derivatives,  for example invariants constructed from the spacetime curvature,  but need not do so in general.
%Second, it may be possible for the symmetry  protecting the moduli space to be a true exact, non-accidental, global internal symmetry, but an axionic one.  Axionic  symmetries cleanly avoid the no-go result of \cite{Banks:1988yz} and are not directly constrained by  the black hole evaporation argument.

\vskip 4pt
\noindent
{\it Existence and genericity of moduli spaces.}---Let us  briefly indicate when exact quantum moduli spaces are generic or non-generic.

In  ${\cal N}=1$ supersymmetry in  Minkowski space, the moduli space is entirely determined by $W$,
which is not renormalized in perturbation theory.  However, nonperturbative effects can contribute corrections to $W$,
lifting the  continuous moduli space  and leaving only discrete
points as quantum vacua.  In ${\cal N}=1$ supergravity theories arising from compactifications of string theory to Minkowski space, the possible nonperturbative effects (from strong  gauge dynamics and from  Euclidean branes)  are numerous,  and  the quantum moduli space is expected to be a set of points  in generic cases.
On the other hand, in {\it{global}} ${\cal N}=1$ supersymmetry in $M_4$, there are celebrated examples  of supersymmetric gauge theories with exact quantum moduli spaces, e.g.~the theory with gauge group $SU(N_c)$ and $N_f=N_c$ families of quarks and anti-quarks in the fundamental representation \cite{Intriligator:1995au}.
%Because the gauge theory is asymptotically free, it is consistent to consider only the two-derivative action.
%In  models with local supersymmetry, in contrast, the  ultraviolet divergences due to gravity oblige us to consider the possibility of higher-derivative terms,  for example from  integrating out massive string states.

In $AdS_4$ the situation changes due to the presence of the K\"ahler potential $K$
in the condition for a supersymmetric minimum. The K\"ahler potential is renormalized
at all orders in perturbation theory, and thus  even perturbative moduli spaces are non-generic.
%one expects perturbative moduli spaces generically only for true shift symmetries of the Lagrangian as in the first example above.
This intuition is supported by the AdS/CFT correspondence, since the three-dimensional SCFT on the boundary of $AdS_4$ only has two supercharges (or four superconformal charges), and thus no BPS representations protected by non-renormalization
theorems exist.

Exact moduli spaces of Minkowski solutions are more common in theories with  extended supersymmetry: for example, in global ${\cal N}=2$ supersymmetry in Minkowski space, the vector multiplet sector
generically
has a quantum moduli space \cite{Lerche:1996xu}.
More generally, even in local supersymmetry, ungauged ${\cal N}=2$ supergravities, for example those arising in the low-energy limit of Calabi-Yau compactifications of type II  string theory, generically
have an exact moduli space in the Minkowski vacuum. The reason for this is that there is no superpotential that can get corrected: only kinetic terms receive quantum corrections. When there are no gaugings, there are no prepotentials, and therefore no potential. On the other hand, when there are gaugings, quantum corrections will correct the potential (since they correct the special K\"ahler and quaternionic-K\"ahler metrics).
For AdS vacua, there must be gaugings, and one expects corrections to the potential.
%  as well as higher-derivative corrections.

\section{${\cal N}=2$ Supergravity in $AdS_4$}\label{sec:N2}
\subsection{Preliminaries}

Let us start with a brief summary of ${\cal N}=2$ supergravity in four space-time dimensions.\footnote{For a more comprehensive
  review see e.g.~\cite{Andrianopoli:1996cm}.}
Apart from the gravitational multiplet,
a generic ${\cal N}=2$ spectrum contains
$n_{\rm v}$ vector multiplets and $n_{\rm h}$ hypermultiplets
with the following field content.
A vector multiplet contains a vector
$A_\mu$, two gaugini $\lambda^{\cal A},{\cal A}=1,2$ and a complex scalar $t$,
while
a hypermultiplet contains two hyperini $\zeta_\alpha$ and four real scalars
$q^u$. Finally, the gravitational multiplet contains the
spacetime metric $g_{\mu \nu}$,
two gravitini $\Psi_{\mu {\cal A}}$ and the graviphoton $A^0_\mu$.\footnote{Strictly speaking, the definition of the graviphoton is $X^I \Im {\cal F}_{IJ} A^J_\mu$, which can be read off from the gravitino variation and depends on the scalar fields in the vector multiplets.}

The scalar field space splits into the product
\begin{equation}\label{Fspace}
M = M_{\rm v}\times M_{\rm h} \ ,
\end{equation}
where the first component $M_{\rm v}$ is a special K\"ahler manifold of complex
dimension $n_{\rm v}$ spanned by the scalars $t^i, i=1,\dots,n_{\rm v}$ in the vector multiplets. This implies that the metric obeys
\begin{equation}\label{Kdef}
g_{i\bar \jmath} = \partial_i \partial_{\bar \jmath} K^{\rm v}\ ,
\qquad \textrm{with}\qquad
K^{\rm v}= -\ln \iu \left( \bar X^\Lambda \Omega_{\Lambda \Sigma}
   X^\Sigma \right)\ ,
\end{equation}
where $X^\Lambda = (X^I, {\cal F}_I), I=0,\ldots,n_{\rm  v}$ is a
$2(n_{\rm  v}+1)$-dimensional symplectic vector that
depends holomorphically on the $t^i$.
${\cal F}_I = \partial{\cal F}/\partial X^I $ is the derivative of a
holomorphic prepotential ${\cal F}$ which is homogeneous of degree 2
and $\Omega_{\Lambda \Sigma}$ is the standard symplectic metric.

The second factor of the field space, $M_{\rm h}$, is spanned by the real scalars $q^u, u=1,\dots, 4n_{\rm h}$ in the hypermultiplets,
and is quaternionic K\"ahler and of real dimension $4n_{\rm h}$.  Such a
manifold admits a triplet of almost complex structures $I^x, x=1,2,3$ satisfying
$I^x I^y = - \delta^{xy} {\bf 1} + \epsilon^{xyz} I^z$, with
the metric $h_{uv}$ being Hermitian with respect to all three $I^x$. The
associated two-forms $K^x$
are the field strengths of the $SU(2)$ connection $\omega^x$, i.e.\
\begin{equation}\label{Kx}
 K^x = \diff \omega^x + \tfrac12 \epsilon^{xyz} \omega^y \wedge \omega^z \ ,
\end{equation}
and thus are covariantly closed, $\nabla K^x=0$.

One of the differences compared to ${\cal N}=1$ supergravity is that no superpotential
is allowed in ${\cal N}=2$ , and thus for Abelian vectors and neutral hypermultiplets
no potential is possible: the entire field space \eqref{Fspace}
is the moduli space of an $M_4$ background.
A potential only appears when some of the hypermultiplets are charged
and/or  when the gauge group is non-Abelian. Let us first discuss
a non-Abelian gauge group $G$.
In this case the scalars $t^i$ are in the adjoint representation
of $G$,  and the contribution to the potential is
%positive
nonnegative and vanishes
for $t^i=0$. Thus a spontaneous breaking of $G$ by a non-trivial
$\langle t^i\rangle$
can
induce a positive contribution
to the  cosmological constant, but cannot be responsible for
the $AdS$ background in the first place. For that reason we discard
non-Abelian gauge groups in the following analysis and only consider
hypermultiplets  that are charged with respect to some Abelian $G= [U(1)]^{n_{\rm _v}}$.
However we do allow for the possibility that the hypermultiplets carry
mutually local electric and magnetic charges.
This situation is conveniently discussed in the embedding tensor
formalism, where the covariant derivatives are given by  \cite{deWit:2005ub}
\begin{equation}\label{Dcov}
 D_\mu q^u = \partial_\mu q^u - A_\mu^\Lambda \Theta_\Lambda^\lambda k^u_\lambda (q)\ ,
\end{equation}
with $A_\mu^\Lambda= (A^I_\mu, B_{\mu\,I})$ being a symplectic vector
of electric and magnetic gauge fields.
Here $k^u_\lambda(q)$
are the independent
Killing vectors on $M_{\rm h}$,  labeled by the
index $\lambda$,  while
$\Theta_\Lambda^\lambda$ is the (constant) matrix of gauge charges
(or the embedding tensor) that parameterizes
the isometries  that are gauged.
Mutual locality additionally imposes the quadratic constraint
$\Theta^{\Lambda\lambda}\Theta_{\Lambda}^\kappa=0$.

The  resulting scalar potential  reads
\begin{equation} \label{eq:N=2potential}
 V = \tfrac12 g_{i \bar \jmath} W^{i{\cal AB}} W^{\bar \jmath}_{\cal
   AB} + N_\alpha^{\cal A} N^\alpha_{\cal A}  - 6 S_{\cal AB} \bar S^{\cal AB} \ ,
\end{equation}
where $W^{i{\cal AB}}, N^\alpha_{\cal A}$ and  $S_{\cal AB}$ arise as the
scalar parts of
 the supersymmetry variations of the gaugino, hyperino and gravitino, respectively
\begin{equation}\label{eq:susyvar}\begin{aligned}
\delta_\epsilon \lambda^{i {\cal A}} &= W^{i{\cal AB}}\epsilon_{\cal B}+\ldots \ ,\\
\delta_\epsilon \zeta_{\alpha} &= N_\alpha^{\cal A} \epsilon_{\cal A}+\ldots \ .\\
 \delta_\epsilon \Psi_{\mu {\cal A}} &= D_\mu \epsilon^*_{\cal A} - S_{\cal AB} \gamma_\mu \epsilon^{\cal B} + \ldots \ .
\end{aligned}\end{equation}
Here $\epsilon^{\cal A}$ are the two supersymmetry parameters,
and
\begin{equation}\label{eq:matrices}\begin{aligned}
S_{\cal AB} =&\, \tfrac{1}{2} \e^{K^{\rm v}/2} X^\Lambda \Theta_\Lambda^{\lambda} P_{\lambda}^x
(\sigma^x)_{\cal AB} \ , \\
W^{i{\cal AB}}
=&\, \mathrm{i} \e^{K^{\rm v}/2} g^{i\bar \jmath}\,
(\nabla_{\bar \jmath}\bar X^\Lambda) \Theta_\Lambda^{\lambda} P_{\lambda}^x (\sigma^x)^{\cal AB}
\ , \\
N_\alpha^{\cal A}
=&\, 2 \e^{K^{\rm v}/2} \bar X^\Lambda \Theta_\Lambda^{\lambda} {\cal U}^{\cal A}_{\alpha u} k^u_{\lambda}
\ ,
\end{aligned}\end{equation}
where in our conventions the Pauli matrices with both indices
up (or down) are\footnote{The indices are raised and lowered with
$\epsilon_{{\cal A}{\cal B}}$.}
 \begin{equation}\label{sigmadef}
(\sigma^1)^{\cal AB} =  \left(\begin{array}{cc}-1 &0\\ 0& 1
   \end{array}\right)~,
\quad
(\sigma^2)^{\cal AB} =  \left(\begin{array}{cc} -\mathrm{i} &0\\ 0&
      -\mathrm{i} \end{array}\right)~ ,
\quad
(\sigma^3)^{\cal AB} =  \left(\begin{array}{cc}0 &1\\ 1& 0 \end{array}\right)~.
\end{equation}
${\cal U}^{\cal A \alpha}_u$ are the vielbeins on $M_{\rm  h}$, which are
related to the metric $h_{uv}$ and the three curvature two-forms
$K^x_{uv}$
defined in \eqref{Kx} via
\begin{equation}\label{Udecomp}
C_{\alpha\beta}{\mathcal U}_{u}^{\mathcal A\alpha }{\mathcal U}_v^{\mathcal B\beta}
= - \tfrac{\iu}{2} K^x_{uv}\sigma^{x {\mathcal A}{\mathcal B}} - \tfrac12 h_{uv}
\epsilon^{{\mathcal A}{\mathcal B}}\ ,
\end{equation}
where $C_{\alpha \beta}, \alpha, \beta=1,\ldots,2n_{\rm h}$ is the flat $Sp(n_{\rm h})$ metric.
Furthermore, we abbreviate
$\nabla_{i}X^\Lambda := \partial_i X^\Lambda + (\partial_i K^{\rm
  v})X^\Lambda$, and
 $P_{\lambda}^x$ are the Killing prepotentials defined by
\begin{equation}\label{eq:Pdef}
-2 k^u_\lambda\,K_{uv}^x  =   \nabla_v P_\lambda^x\ ,
\end{equation}
where $\nabla_v$ is the $SU(2)$-covariant derivative.

In the following discussion we will also need the fermion mass matrices \cite{Andrianopoli:1996cm}
\begin{equation}\label{eq:fermionmassmatrices}\begin{aligned}
{\cal M}_{ij \cal AB} = &  \tfrac12  \e^{K^{\rm v}/2} (\nabla_i \nabla_j X^\Lambda) \Theta_\Lambda^{\lambda} P_{\lambda}^x
(\sigma^x)_{\cal AB}\ , \\
{\cal M}^\alpha_{i \cal A} = & - 4 \e^{K^{\rm v}/2} (\nabla_i X^\Lambda) \Theta_\Lambda^{\lambda} {\cal U}^{\alpha}_{{ \cal A}u} k^u_\lambda \ , \\
{\cal M}^{\alpha \beta} = &  - \e^{K^{\rm v}/2} X^\Lambda  \Theta_\Lambda^{\lambda} {\cal U}^{\alpha \cal A}_u {\cal U}^{\beta \cal B}_v \epsilon_{\cal AB} \nabla^{[u} k^{v]}_\lambda  \ ,
\end{aligned} \end{equation}
where ${\cal M}_{ij \cal AB}$ is the mass matrix of the gauginos,
$ {\cal M}^{\alpha \beta}$ is the mass matrix of the hyperini and
${\cal M}^\alpha_{i \cal A}$ is a possible mixing term.\footnote{Strictly speaking the  fermion mass matrices are the values of these quantities evaluated in the AdS background.}
Supersymmetry relates the shift matrices in \eqref{eq:matrices} and
the fermion mass matrices
\eqref{eq:fermionmassmatrices} by the following ``gradient flow'' equations \cite{D'Auria:2001kv}
\begin{equation} \begin{aligned}
 \nabla_j W^{i\cal AB} &=  2 \delta^i_j \bar S^{\cal AB} \ , \qquad
 \nabla_{\bar \jmath} W^{i\cal AB} =  - g^{i\bar \imath} {\cal M}^{\cal AB}_{\bar \imath \bar \jmath} \ , \qquad
 \nabla_u \bar W^{\bar \jmath}_{\cal AB} =  - \tfrac12 g^{i \bar \jmath} {\cal M}^\alpha_{i (\cal A} {\cal U}_{{\cal B})\alpha u} \ , \\
 \nabla_i N^\alpha_{\cal A} &= \tfrac12 {\cal M}^\alpha_{i \cal A} \ , \qquad
 {\cal U}^{u \beta \cal B} \nabla_u N^{\alpha \cal A} =  4 C^{\alpha \beta} S^{\cal AB} + \epsilon^{\cal AB} {\cal M}^{\alpha \beta} \ , \\
 \nabla_i S_{\cal AB} &=  \tfrac12 g_{i\bar \jmath} W^{\bar \jmath}_{\cal AB} \ ,\qquad
 \nabla_{\bar \imath} S_{\cal AB} =  0 \ ,  \qquad
\nabla_u S_{\cal AB} =  -\tfrac12 U_{u \alpha (\cal A} N^\alpha_{\cal B)} \ ,
\end{aligned}\end{equation}
where $W^{\bar \jmath}_{\cal AB} = (W^{i\cal AB})^*$ and ${\cal M}^{\cal AB}_{\bar \imath \bar \jmath} = ({\cal M}_{ij \cal AB})^*$.

\subsection{Structure of the moduli space}

In \cite{Hristov:2009uj,Louis:2012ux} the conditions for a
four-dimensional ${\cal N}=2$ supersymmetric AdS vacuum
in ${\cal N}=2$ supergravity were discussed.
In terms of the fermionic supersymmetry variations \eqref{eq:susyvar}
one demands
\begin{equation}\label{eq:susy}
 \vev{W^{i\cal AB}} = 0 \ , \qquad \vev{N^{\alpha \cal A}} = 0 \ ,\qquad
\vev{S_{\cal AB}} \epsilon^{\cal B} = \tfrac12 \Lambda \epsilon^*_{\cal A}\ ,
\end{equation}
where $|\Lambda|^2$ is related to the cosmological constant of the
${\cal N}=2$ vacuum as in \eqref{Lambdarel}.
Using \eqref{eq:matrices} the conditions \eqref{eq:susy}
can be explicitly translated into the following conditions on the ${\cal N}=2$ couplings \cite{Louis:2012ux}
\begin{equation}\label{eq:hyperino}
 \vev{X^\Lambda \Theta_\Lambda^\lambda k_\lambda^u }= 0 \ ,
\qquad \vev{\nabla_i X^\Lambda \Theta^\lambda_\Lambda P^x_{\lambda}} =
0\ ,
\end{equation}
and
\begin{equation}\label{eq:gaugino}
\Theta_\Lambda^{\, \lambda}\, \vev{P^x_\lambda} = - \tfrac12
 \Omega_{\Lambda \Sigma}
\vev{\e^{K^{\rm v}/2}  \Im (\hat \Lambda \bar X^\Sigma)}\, a^x  \ ,
\end{equation}
where $a$ is an arbitrary real vector on $S^2$
and $\hat \Lambda$ is related to $\Lambda$ by a phase.
We can use the local $Sp(1)$ symmetry of ${\cal N}=2$ to rotate $a^x$
into a frame where  $a^x = a \delta^{x3}$ and hence only the combination
$\Theta_\Lambda^{\, \lambda}  \vev{P^3_\lambda}\neq 0$ in
\eqref{eq:gaugino}.
(We will frequently
use this simplification below.)

By contracting \eqref{eq:gaugino} with $X^\Lambda$ it was shown in
\cite{Louis:2012ux} that the right hand side
 is proportional to the graviphoton direction
in field space, and thus a cosmological constant can only appear if
an isometry in this direction is gauged. Let us denote this direction by
$\lambda=0$, so that
\eqref{eq:gaugino} implies
\begin{equation}\label{eq:Ps}
\vev{X^\Lambda\Theta^0_\Lambda P^x_0}\neq 0\ ,\qquad
\vev{X^\Lambda\Theta^{\lambda\neq0}_\Lambda P^x_{\lambda\neq0}} = 0 \ ,
\end{equation}
and thus, inserted into \eqref{eq:matrices},
\beq\label{Ssimp}
\vev{S_{\cal AB}} =\, \tfrac{1}{2} \vev{\e^{K^{\rm v}/2} X^\Lambda \Theta_\Lambda^{0} P_{0}^x}\,
(\sigma^x)_{\cal AB} =\, \tfrac{1}{2} \vev{\e^{K^{\rm v}/2} X^\Lambda \Theta_\Lambda^{0} P_{0}^3}\,
(\sigma^3)_{\cal AB} \neq0\ .
\eeq
The first equation in \eqref{eq:hyperino} combined with the requirement
\eqref{eq:Ps} has two types of solutions:
\beq\begin{aligned}\label{solutions}
\textrm{minimal solution}:
\qquad& \langle k_\lambda^u\rangle =0\quad \forall\, \lambda \ ,\\
\textrm{non-minimal solution}: \qquad& \langle k_0^u\rangle =0\ ,\quad
\langle k_{\lambda\neq0}^u\rangle \neq0\ .
\end{aligned}\eeq
For the
non-minimal solution, \eqref{eq:hyperino} is satisfied only by imposing
$\vev{X^\Lambda\Theta^{\lambda\neq0}_\Lambda} = 0$.
In this case the gauge symmetry is spontaneously broken
and  $n_m:={\rm rk}(\Theta_\Lambda^\lambda k_\lambda^u)$
long
vector multiplets become massive,
%\lm{"each" added}
each with a  total of five massive scalars,
two from vector multiplets and three from
hypermultiplets~\cite{Louis:2012ux}.\footnote{The fourth hyper-scalar
  is the Goldstone boson eaten by the vector.}
Note that consistency imposes
\beq
n_m\le n_v\qquad \textrm{and}\qquad
n_m\le n_h\ .
\eeq

As in section \ref{sec:N1}, we now determine properties of the moduli
space by varying the conditions \eqref{eq:hyperino} and
\eqref{eq:Ps}. Let us start with \eqref{eq:Ps} and study the variation
 of $\vev{X^\Lambda\Theta^\lambda_\Lambda P^x_\lambda}$ for all $\lambda$.  This has the two
terms
\begin{equation}
 \vev{\delta (X^\Lambda\Theta^\lambda_\Lambda P^x_\lambda)} =  \vev{\nabla_i X^\Lambda \Theta^\lambda_\Lambda P^x_\lambda}\, \delta t^i  + 2  \vev{X^\Lambda \Theta_\Lambda^\lambda k_\lambda^v K^x_{uv}}\, \delta q^u  =
0 \ ,
\end{equation}
which both vanish for both solutions of \eqref{solutions},
due to \eqref{eq:hyperino}. Thus no condition is imposed on the moduli space.

Next we consider
the variation of
$\vev{\nabla_i X^\Lambda \Theta^\lambda_\Lambda  P^x_\lambda} $
in \eqref{eq:hyperino}, which
yields
\begin{equation} \label{eq:N=2SKmoduli}
\vev{{\cal M}_{ij \cal AB}}\, \delta t^j - 2 \vev{\bar S_{\cal AB}\, g_{i\bar \jmath} }\, \delta \bar t^{\bar \jmath}  + \tfrac12 \vev{M^\alpha_{i (\cal A}{\cal U}_{{\cal B})\alpha u}}\, \delta q^u = 0 \ ,
\end{equation}
where ${\cal M}_{ij\, \cal AB}$ and $M^\alpha_{i \cal A}$
are  defined in \eqref{eq:fermionmassmatrices} and we used \eqref{eq:matrices}
and \eqref{eq:Pdef}.
For the minimal solution we find from the definition in
\eqref{eq:fermionmassmatrices} that the mass matrix ${\cal
  M}^\alpha_{i \cal A}$ vanishes
and one is left with only the first two terms in \eqref{eq:N=2SKmoduli}.
As we anticipated the analysis can be
further simplified by using
the local $Sp(1)$ symmetry of ${\cal N}=2$ to rotate
into a frame where among the $\Theta^\lambda_\Lambda P^x_\lambda$ only $\Theta^\lambda_\Lambda P^3_\lambda$ is non-zero.
Inserting \eqref{eq:fermionmassmatrices} and \eqref{eq:gaugino} into
\eqref{eq:N=2SKmoduli} in that frame
yields
\begin{equation} \begin{aligned}\label{eq:N=2SKmodulimixs}
\vev{  \Im (\hat \Lambda \bar X^\Sigma)}\Omega_{ \Sigma \Lambda}
\big(\vev{ \nabla_i \nabla_j
X^\Lambda } \delta t^j - 2 \vev{ X^\Lambda
g_{i\bar\jmath}} \delta \bar t^{\bar \jmath}\big) = 0 \ .
\end{aligned}\end{equation}
These are $n_v$ complex equations, and
comparing with \eqref{DW=0sol} we see that
the ${\cal N}=1$
%\lm{typo fixed, "=1" was missing}
analysis of the previous section applies verbatim. Thus the
$AdS_4$ moduli space of the vector multiplets is again real and at most of
(real) dimension $n_v$.

Let us postpone the discussion of \eqref{eq:N=2SKmoduli} for the
non-minimal solution where $\vev{{\cal M}^\beta_{i \cal A} }\neq0$ and instead
turn to the first condition in  \eqref{eq:hyperino}.
The variation of $\vev{X^\Lambda \Theta_\Lambda^\lambda
  k_\lambda^u }$ yields
\begin{equation}\label{N2var}
\tfrac12 C_{\alpha \beta} \vev{{\cal M}^\beta_{i \cal A} }\, \delta t^i   + \vev{ \mathbb{M}_{{\cal A}\alpha {\cal B}\beta}{\cal U}^{{\cal B}\beta}_u}\, \delta q^u   = 0 \ ,
\end{equation}
where we defined
\begin{equation}\label{masshyper}
\mathbb{M}_{{\cal A}\alpha {\cal B}\beta} =
4\e^{K_{\rm v}/2} X^\Lambda \Theta_\Lambda^\lambda (\nabla_v k_{\lambda u})\,{\cal U}^v_{{\cal B}\beta}{\cal U}_{{\cal A}\alpha}^u =
4 C_{\alpha \beta} S_{\cal AB} -  \epsilon_{\cal AB} M_{\alpha\beta} \ ,
\end{equation}
and the last equality used \eqref{eq:matrices} and
\eqref{eq:fermionmassmatrices}. As before let us first analyze
the minimal solution with $\vev{{\cal M}^\beta_{i \cal A} }=0$ and the first term in \eqref{N2var} vanishing.
Due to \eqref{eq:Ps} we rotate into the frame where only
\begin{equation}
P^3_0 = 2 \vev{ \e^{K_{\rm v}/2} X^\Lambda \Theta^\lambda_\Lambda P^3_\lambda} = \vev{\e^{K_{\rm v}} \Im (\hat \Lambda \bar X^\Sigma) \Omega_{\Sigma \Lambda} X^\Lambda} \ ,
\end{equation}
is non-zero, and
using \eqref{sigmadef} and \eqref{Ssimp} the $(4n_{\rm h}\times
4n_{\rm h})$ matrix
$\mathbb{M}_{{\cal A}\alpha {\cal B}\beta}$ then takes the form
\begin{equation}
\mathbb{M}_{{\cal A}\alpha {\cal B}\beta} =  \left( \begin{array}{cc} 0 & C_{\alpha \beta}P^3_0 - M_{\alpha\beta} \\ C_{\alpha \beta}P^3_0 + M_{\alpha\beta} & 0 \end{array} \right) \ .
\end{equation}
Note that $C_{\alpha \beta}$ is anti-symmetric while $M_{\alpha\beta}$
is symmetric, and as a consequence ${\mathbb{M}}^\top
= -\mathbb{M}$. That is,
$\mathbb{M}$ is altogether antisymmetric, and thus
its eigenvalues come in pairs.
For the case at hand this means
that  the hypermultiplet scalars become  massive pairwise,
and similarly the zero modes come in pairs.
Furthermore, $C$ is the flat $Sp(n_{\rm h})$ metric and thus by
appropriately tuning $M_{\alpha\beta}$ one can reduce the rank
of both $(2n_{\rm h}\times 2n_{\rm h})$
matrices $C_{\alpha \beta}P^3_0 \pm M_{\alpha\beta}$ to be
$n_{\rm h}$, but no smaller. In other words
$\mathbb{M}_{{\cal A}\alpha {\cal B}\beta}$ has at least rank $2n_{\rm h}$
(instead of $4n_{\rm h}$),
and at most $2n_{\rm h}$ scalars can be massless.

Let us now show that the ${\cal N}=2$ $AdS_4$ moduli space of the
hypermultiplet scalars is K\"ahler.
First of all there is a complex structure given by $I^3$ that acts on
the flat indices as $\sigma^3$.
Indeed it is easy to check that
\begin{equation}\label{eq:sigma3anticomm}
(\sigma^3)^{\cal C}_{\cal A}\, \mathbb{M}_{{\cal C}\alpha {\cal B}\beta} = - \mathbb{M}_{{\cal A}\alpha {\cal C}\beta}\,(\sigma^3)^{\cal C}_{\cal B}\ ,
\end{equation}
so that in particular the massless spectrum
is invariant,
and the moduli space is a complex manifold (with respect to $I^3$).
We will now show that $\vev{K^1} =\vev{K^2}=0$ which via \eqref{Kx}
then implies
\begin{equation}\begin{aligned}
\vev{ \diff K^3} = & - \vev{\omega^1 \wedge K^2} + \vev{\omega^2 \wedge K^1} = 0 \ .
\end{aligned} \end{equation}
Thus $K^3$ is closed on the ${\cal N}=2$ locus, which shows that the $AdS_4$ moduli
space
is not only complex but actually K\"ahler, with $K^3$ as its K\"ahler form. Another consequence of $\vev{K^1} =\vev{K^2}=0$ is that the resulting moduli space is real with respect to the complex structures $I^1$ and $I^2$.

Let us prove $\vev{K^1}=0$ explicitly --- $\vev{K^2}=0$ then follows by
permutations of indices.
Using \eqref{Udecomp} and the algebra of the Pauli matrices we can write $K^1$ as
\begin{equation}\begin{aligned}
 \vev{K^1_{uv}} = & ((\sigma^2)_A^D (\sigma^3)_D^C - (\sigma^3)_A^D (\sigma^2)_D^C)\epsilon_{CB} C_{\alpha \beta} \vev{U^{\alpha A}_u U^{\beta B}_v } \\
 =& ((\sigma^2)_A^C (\sigma^3)_{CB} + (\sigma^3)_{AC} (\sigma^2)_B^C ) C_{\alpha \beta}  \vev{U^{\alpha A}_u U^{\beta B}_v } \\
 =& ((\sigma^2)_A^C \epsilon_{CB}+ \epsilon_{AC} (\sigma^2)_B^C) \vev{M_{\alpha \beta} U^{\alpha A}_u U^{\beta B}_v}  = 0 \ ,
\end{aligned} \end{equation}
where  in the last step we used \eqref{N2var}. This completes our
proof.
We  have shown
that the $AdS_4$ moduli space of the scalars in ${\cal N}=2$ hypermultiplets
is a K\"ahler submanifold of the
parent quaternionic-K\"ahler manifold (which has real dimension $4n_{\rm h}$), and has real dimension at most $2n_{\rm h}$. In fact this coincides
with the mathematical theorem that a K\"ahler submanifold of a
quaternionic-K\"ahler manifold can have at most half the dimension
of the parent \cite{Alekseevsky:2001qk}.

Thus for the minimal solution in \eqref{solutions} the moduli space is a direct product of a real manifold spanned by the vector multiplet scalars and a K\"ahler manifold
spanned by the hypermultiplet scalars.
This is indeed consistent with  the AdS/CFT expectation of a K\"ahler moduli
space for the three-dimensional boundary theory, since
in  three-dimensional supersymmetric theories with four supercharges
the vector multiplet contains as bosonic components a vector and a real scalar.
Dualizing the vector to a real scalar, the entire multiplet becomes dual to a chiral
multiplet. The associated K\"ahler moduli
space can only appear  after dualizing the vector: in the four-dimensional bulk  description
the K\"ahler structure is not visible.
Furthermore, in the minimal solution we have a direct product of moduli spaces,
which is a special case of the generic situation
in three-dimensional supersymmetric theories. As we will see shortly,
this feature will not hold
for the non-minimal solution, and a mixing between vector and hypermultiplet scalars occurs.

As promised let us now discuss the moduli space for the non-minimal
solution in \eqref{solutions} which has $\langle k_0^u\rangle =0,\langle
k_{\lambda\neq0}^u\rangle \neq0, \vev{M^\alpha_{i \cal A}}\neq 0$.
Thus we have to reconsider the variations \eqref{eq:N=2SKmoduli} and
\eqref{N2var}, as both sets of equation have additional terms.
Before we plunge into the technical analysis let us sketch the intuition.
For $\langle k_{\lambda\neq0}^u\rangle \neq0$
 the gauge symmetry is spontaneously broken
and  $n_m$ long vector
multiplets become massive. In this case a vector multiplet eats an
entire hypermultiplet and thus consists of a Goldstone boson from the
hypermultiplet and
 five massive scalars,
two from the vector multiplets and three from the hypermultiplets.
As we will see shortly this situation is
manifest in \eqref{eq:N=2SKmoduli} and \eqref{N2var}, but
the structure of the moduli space is unchanged and only its dimension
is reduced as
$n_m$ additional vector and
hypermultiplets are fixed.

We start by rewriting \eqref{N2var} and \eqref{eq:N=2SKmoduli} in a more explicit form, using \eqref{eq:fermionmassmatrices}, and simplify the
analysis by  rotating into the frame where only $P^3_0$ is non-zero.
This yields
\begin{align}
\label{eq:N=2nonminimal}
& -2  \Theta^\lambda_\Lambda \vev{\e^{K^{\rm v}/2} U_{\alpha {\cal A}u}k_\lambda^u \nabla_i X^\Lambda}\, \delta t^i   + \vev{ \mathbb{M}_{{\cal A}\alpha {\cal B}\beta}{\cal U}^{{\cal B}\beta}_u}\, \delta q^u   = 0 \ ,\\[1ex] \label{eq:N=2SKmoduli12}
&\qquad\Theta^\lambda_\Lambda \langle k^w_\lambda
K^{1,2}_{wu}  \rangle\, \delta q^u  = 0 \ , \\[1ex]
\label{eq:N=2SKmodulimix}
&\vev{\e^{K^{\rm v}/2}  \Im (\hat \Lambda \bar X^\Sigma)}  \Omega_{ \Sigma \Lambda} (\langle \nabla_i \nabla_j
X^\Lambda \rangle \delta t^j - 2 \langle X^\Lambda
g_{i\bar\jmath}\rangle \delta \bar t^{\bar \jmath}) + 4 \langle
\nabla_i X^\Lambda \rangle  \Theta^\lambda_\Lambda \langle k^w_\lambda
K^3_{wu}  \rangle \delta q^u   = 0 \ .
\end{align}

As we already stated, for an Abelian theory the Goldstone bosons
have to be recruited
out of hypermultiplets and thus the $n_m$
Goldstone directions drop out of (\ref{eq:N=2nonminimal})-(\ref{eq:N=2SKmodulimix}). This can be seen by explicitly inserting $\delta q^u\sim  \Theta^\kappa_\Sigma k^u_\kappa$. For  (\ref{eq:N=2SKmoduli12}) and (\ref{eq:N=2SKmodulimix})
we can use the equivariance condition
(see for example \cite{Andrianopoli:1996cm})
\begin{equation} \label{eq:equivariance}
K^x_{uv} k^u_\lambda k^v_\sigma - \tfrac12 \epsilon^{xyz} P^y_\lambda P^z_\sigma = \tfrac12 f^{\rho}_{\lambda\sigma} P^x_\rho \ ,
\end{equation}
where $f^{\rho}_{\lambda\sigma}$ are the structure constants of the gauge algebra, i.e.\
\begin{equation}
[k_\lambda, k_\sigma] = f^{\rho}_{\lambda\sigma} k_\rho \ , \qquad
k_\lambda\equiv k_\lambda^u\partial_u\ .
\end{equation}
Since we only consider Abelian gauged isometries we have
$f^{\rho}_{\lambda\sigma}=0$, and as only $P^3_\lambda$ is non-vanishing, we find from \eqref{eq:equivariance} that
\begin{equation}
K^x_{uv} k^u_\lambda k^v_\sigma = 0 \ .
\end{equation}
This immediately implies that the Goldstone directions $\delta q^u\sim  \Theta^\kappa_\Sigma k^u_\kappa$ indeed drop out of (\ref{eq:N=2SKmoduli12}) and (\ref{eq:N=2SKmodulimix}).
For \eqref{eq:N=2nonminimal} we use \eqref{masshyper}
to find that the Goldstone directions do not contribute as a consequence of
\begin{equation}
 X^\Lambda \Theta^\lambda_\Lambda \Theta^\sigma_\Sigma k_\sigma^v \nabla_v k^u_\lambda =  X^\Lambda \Theta^\lambda_\Lambda \Theta^\sigma_\Sigma [k_\sigma , k_\lambda]^u = 0 \ ,
\end{equation}
where we used  \eqref{eq:hyperino} in the first equality, and in the second equality we used that the gauged directions are Abelian. Thus, the Goldstone directions also drop out of \eqref{eq:N=2nonminimal}. For later use let us  note that due to \eqref{eq:sigma3anticomm} and
$(\sigma^3)^{\cal A}_{\cal B} U^{\alpha {\cal B}}_u
=U^{\alpha {\cal A}}_v (I^3)^v_u$,
the deformations $\delta q^u \sim \Theta^\sigma_\Sigma (I^3)^u_v k_\sigma^v$ also do not appear in \eqref{eq:N=2nonminimal}.

Let us now discuss which scalars are fixed by (\ref{eq:N=2nonminimal})-(\ref{eq:N=2SKmodulimix}).
From \eqref{eq:N=2SKmoduli12}
we immediately see that $2n_m$ hypermultiplet scalars become
massive. They  are related to a Goldstone boson by $I^1$ or $I^2$
and therefore reside in the same hypermultiplet.
We continue with \eqref{eq:N=2nonminimal} and consider first the
 indices $({\cal A}\alpha)$ for which  $U_{\alpha {\cal A}u}k_\lambda^u=0$ holds.
In this case the first term  vanishes  and
 the discussion for the minimal solutions applies.
For indices which have $U_{\alpha {\cal A}u}k_\lambda^u\neq0$ \eqref{eq:N=2nonminimal}
fixes  the $n_m$ complex scalars that are in the same multiplets as the massive vectors.
Finally let us turn to \eqref{eq:N=2SKmodulimix}.
For indices  $i$
that have  ${\cal M}_{i {\cal A}}^\alpha=0$ the last term in
\eqref{eq:N=2SKmodulimix} vanishes and one is left with the minimal case for the vector multiplets that we discussed above.
For indices  $i$
that have  ${\cal M}_{i {\cal A}}^\alpha \neq 0$  the last term fixes $n_m$ additional scalars which are related by $I^3$ to the Goldstone direction.

We have thus shown
that for any gauged non-vanishing Killing vector, $5 n_m$ scalars are massive: these are members of $n_m$ long massive vector multiplets.
Therefore, at least $(n_v + n_m)$ vector multiplet scalars become massive, and the moduli
space has at most real dimension $(n_v - n_m)$. For the hypermultiplets, at least $2 n_h +n_m$ scalars become massive, and $n_m$ scalars are eaten. Thus, there are at most  $(n_h - n_m)$ complex directions corresponding to hypermultiplet moduli.
%\lm{slight rephrasing above and below}
However, compared to the minimal solution both sectors mix non-trivially.
The massive scalars in the long vector multiplets are actually combinations of vector multiplet and hypermultiplet scalars, due to the
fact that the the three mass matrices ${\cal M}_{ij \cal AB}$, $ {\cal M}^{\alpha \beta}$, and ${\cal M}^\alpha_{i \cal A}$ in \eqref{eq:N=2SKmoduli} and \eqref{N2var} are nonvanishing. This in turn
leads to a mixing of vector  multiplet and hypermultiplet scalar fields in the kinetic terms, and thus the moduli space is no longer a direct product.

%%%%%%%%%%%%%%%%%%%%%%%%%%%%%%%%%%%%
\subsection*{Acknowledgements}
%%%%%%%%%%%%%%%%%%%%%%%%%%%%%%%%%%%%

This work was partly supported by the German Science Foundation (DFG) under the Collaborative Research Center (SFB) 676 Particles, Strings and the Early Universe.  We have benefited from conversations and correspondence with
Micha Berkooz,
Vicente Cort\'es, Oliver DeWolfe,  Jerome Gauntlett, Shamit Kachru, Paul Smyth,
Dan Waldram and Marco Zagermann.
J.~L.\ and L.~M.\ thank the Simons Center and the organizers of
the 2012 Summer School on String Phenomenology,
where this project was initiated, for their hospitality. S.~dA.,  L.~M.~and H.~T.~thank the organizers of String Phenomenology 2013  for providing a stimulating environment for the completion of this work.
The research of L.~M.~was  supported in part by the NSF under grant PHY-0757868. S. dA. would like to acknowledge the award of a CRC 676 fellowship from DESY/Hamburg University and a visiting professorship at Abdus Salam ICTP. His research was supported in part  by the United States Department of Energy under grant DE-FG02-91-ER-40672.
H.~T.~thanks the IPhT at CEA Saclay, where part of this work was performed. His work was supported in part by the ANR grant 08-JCJC-0001-0 and the ERC Starting Grants 259133 -- ObservableString and 240210 - String-QCD-BH. A.~W. was supported by the Impuls und Vernetzungsfond of the Helmholtz Association of German Research Centres under grant HZ-NG-603.

\newpage

\appendix
\noindent
{\bf\Large Appendix}

\section{Relation to Global ${\cal N}=1$ Supersymmetry in $AdS_4$}  \label{globalAdS}

In this appendix we recall the global limit of a generic supergravity theory in
$AdS_4$. For this we need to restore the dependence on the gravitational coupling $\kappa \equiv \Mp^{-1}$.
We take all fields to have mass dimension one
(denoted by $[\phi]=1$), and correspondingly take $[K]=2$, $[W]=3$, and $[V]=4$.
In these conventions the potential \eqref{eq:V1} reads
\begin{equation}\label{eq:V}
V=e^{\kappa^2K}\Bigl(K^{i\jbar}D_{i}W D_{\jbar}\bar{W}-3\kappa^2|W|^{2}\Bigr)\ ,\qquad
\textrm{with}\qquad D_{i}W= \partial_i W + \kappa^2 K_i W\ .
\end{equation}
Without loss of generality we can parameterize  the superpotential as
\beq
W= W_0 + \P\ ,
\eeq with
\beq
\langle \P \rangle=0\ \qquad {\rm{and}} \qquad  \kappa^2 W_0 \equiv \Lambda \ne 0\ ,
\eeq  where we have taken $W_0$  to be real.  Furthermore, by a  choice of  K\"ahler gauge  we may set $\langle K \rangle=0$.
The cosmological constant $\kappa^2 \langle V\rangle$ that appears in the Einstein equations
is related to $\Lambda$ by
\beq\label{Lambdarel}
\kappa^2 \langle V\rangle=-3 \Lambda^2\ .
\eeq
In order to obtain an $AdS_4$ background for global supersymmetry from supergravity,
one needs to take the limit
$\kappa\to 0, \Lambda$ fixed.
Expanding $V$ in this limit one arrives at\footnote{Note that $V$
diverges in this limit, but the Einstein equations are finite.} \cite{Adams:2011vw}
\beq
V = K^{i\bar{\jmath}} {\cal{D}}_i \bar{{\cal{D}}}_{\bar \jmath} -3\Lambda
 (\P+\bar\P+\Lambda K)
-3\kappa^{-2}\Lambda^{2}+{\cal O}(\kappa^2)\ ,
\label{eq:VAdS}
\eeq
where ${\cal{D}}_i  \equiv \partial_{i}W + K_{i}\Lambda$ (not to be confused with $D_i \equiv \partial_i + K_i$)
vanishes at the
supersymmetric minimum, i.e.\
$\langle {\cal{D}}_i \rangle=0$.
The first derivative of $V$ reads
\beq
\partial_k V= \nabla_k V = K^{i\bar{\jmath}} (\nabla_k {\cal{D}}_i) \bar{{\cal{D}}}_{\bar{\jmath}}-2\Lambda {\cal{D}}_k +{\cal O}(\kappa^2)\ ,
\label{eq:V1AdS}
\eeq
which indeed vanishes at the minimum, because $\langle {\cal{D}}_i \rangle=0$.

From \eqref{eq:V1AdS} we can compute the `mass matrix'
\beq\begin{aligned}
\langle \nabla_k \nabla_{\bar l} V\rangle &=
-2 K_{k\bar l} \Lambda^2 + K^{i\bar{\jmath}}
m_{ki}\bar{m}_{\bar l\bar{\jmath}}\\
\langle \nabla_k \nabla_{l} V\rangle &=-m_{kl}\Lambda
\end{aligned}
\eeq
where
\begin{equation} \label{fermionmass}
m_{ki}= e^{\kappa^2 K/2}\nabla_k {\cal{D}}_i
\end{equation}
is the fermionic mass matrix.
Decomposing $\phi^i = \frac1{\sqrt 2}(A^i+iB^i)$ we obtain the
mass matrices for $A^i$ and $B^i$,
\beq\begin{aligned}  \label{decoupledmassmatrix}
(m_{A}^{2})_{kl} & = K^{i\bar{\jmath}} m_{ki}\bar{m}_{\bar l\bar{\jmath}}-\Lambda
m_{kl}
-2\Lambda^{2}K_{k\bar l}\ ,\\
(m_{B}^{2})_{kl} & = K^{i\bar{\jmath}} m_{ki}\bar{m}_{\bar l\bar{\jmath}}+\Lambda
m_{kl}
-2\Lambda^{2}K_{k\bar l}\ ,\\
(m^2_{AB})_{kl}& =2\,\Im m_{kl}\Lambda \ .
\end{aligned}\eeq
The mass matrices (\ref{decoupledmassmatrix})  agree with \cite{deWit:1999ui} when there is only one chiral multiplet and  $\Im m$ is taken to be zero. For $\Im m \neq 0$ we can consider for simplicity the case of one multiplet with canonical K\"ahler metric. In this case the matrices are easily diagonalized, with the mass-squared eigenvalues
 \beq\label{physmass}
M_{\pm}^{2}= |m|^2\pm\Lambda
|m|
-2\Lambda^{2}.
\eeq
We see that one cannot have $M_+=M_-=0$ without setting $\Lambda =0$.

One might be tempted to think that  flat directions in the potential arise when $m_{ij}=0$, as in flat space.
This is incorrect,  as we now explain in the simple case of a single chiral multiplet scalar $\phi$ (so that $m_{ij} \rightarrow m$), with $K=\phi{\bar\phi}$.
The equation of motion in Einstein frame reads
\beq\label{force}
\nabla^{2}\phi=\frac{1}{\sqrt{g}}\partial_{\mu}\sqrt{g}g^{\mu\nu}\partial_{\nu}\phi=V_{\bar\phi}\quad,
\eeq
and for $m=0$ the right hand side does not vanish, cf.~(\ref{eq:V1AdS}).
In fact, a scalar field $\phi$ with $m=0$ is a
{\it{conformally coupled}} scalar.
To see this, we  perform a Weyl rescaling to the Jordan frame, so that
the Lagrangian reads
%\Jnote{check}
\beq
{\cal L} =
-\frac{1}{2\kappa^{2}}R\, e^{-\frac{\kappa^2 }{3}K}
-K^J(\phi,\bar{\phi})\,\partial_{\mu}\phi\partial^{\mu}\bar{\phi}-V^{J}(\phi,\bar{\phi})+\ldots\ ,
\label{eq:Jaction}
\eeq
where $K^J, V^J$ are the metric and potential in the Jordan frame.
In the limit $\kappa\to 0, \Lambda$ fixed one finds
\beq
V^J = K^{i\bar{\jmath}}{\cal D}_i {\bar{\cal D}}_{\bar{\jmath}} -3\Lambda
 (\P+\bar\P)-\Lambda^2 K
-3\kappa^{-2}\Lambda^{2}+{\cal O}(\kappa^2)\ ,
\label{eq:VJ}
\eeq
where compared to \eqref{eq:VAdS} only the coefficient
of the $\Lambda^2 K$ term has changed.
As a consequence, the  mass matrix \eqref{physmass} in the Jordan frame takes the form
 \beq\label{MJ}
M_{J \pm}^{2}= |m|(|m|\pm\Lambda)
\eeq
which vanishes for $m=0$.
Moreover, the equation of motion in the Jordan frame becomes
\be
 \nabla^{2}\phi=V_{\bar{\phi}}^{J}-\frac{R}{6}\phi+{\cal O}(\kappa^{2})=V_{\bar{\phi}}^{J}-2\Lambda^{2}\phi+{\cal O}(\kappa^{2}) = V_{\bar{\phi}},\ee
 where in the second equality the Einstein equations have been used, and the final relation uses \eqref{eq:VAdS}.
Thus, $m=0$  implies that $V_{\bar{\phi}}^{J}=0$, but that (for $\phi \neq 0$) $V_{\bar{\phi}} \neq 0$, and hence $\nabla^{2}\phi \neq 0$.
In summary, a field with $m=0$ is conformally coupled, but is not a modulus.

\begin{figure}[t]
\begin{center}
\includegraphics[width=0.95\textwidth]{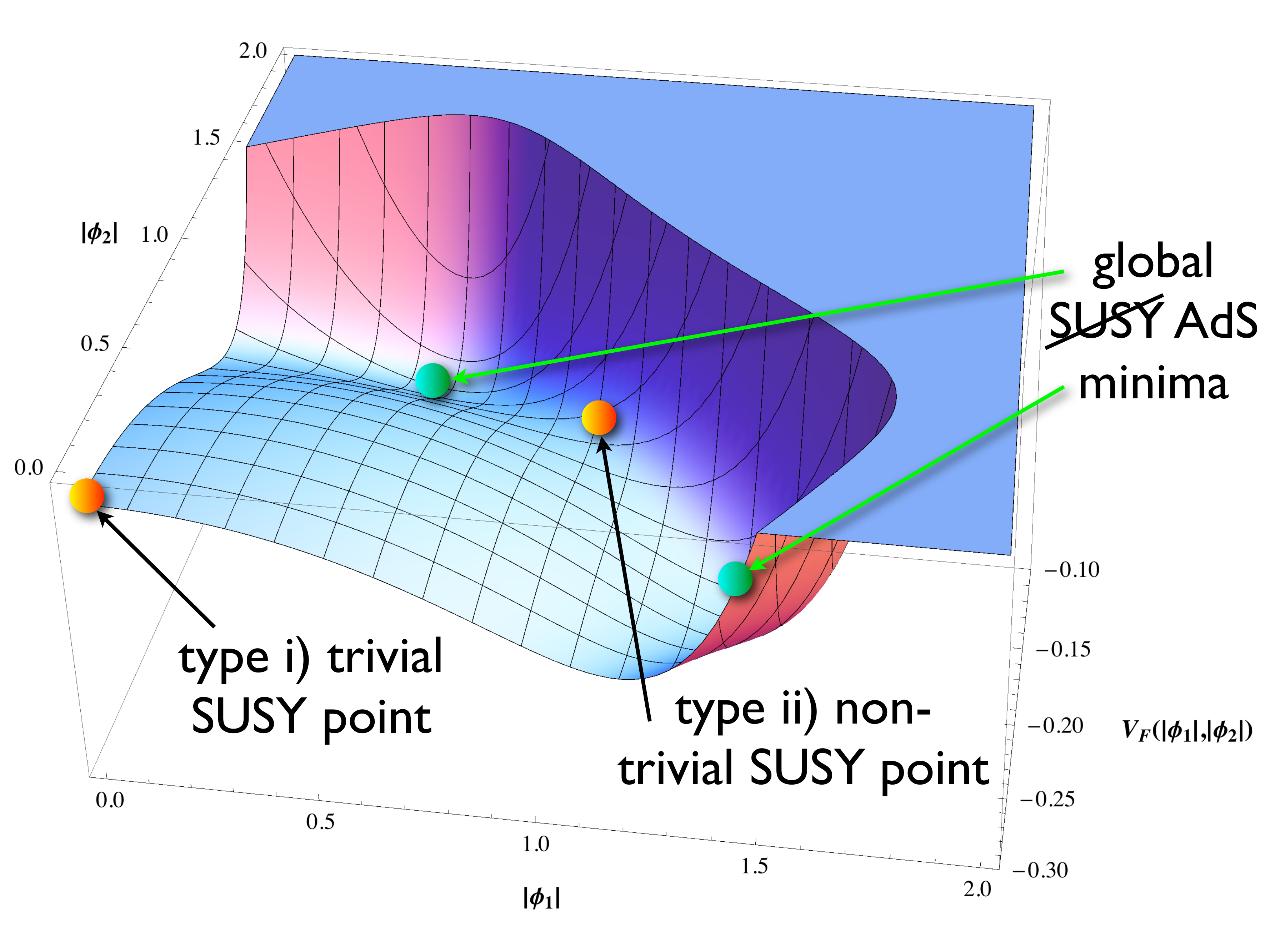}
\end{center}
\refstepcounter{figure}\label{Fig.1}
\vspace*{-.2cm} {\bf Figure~\ref{Fig.1}:} The potential $V(|\phi_1|,|\phi_2|)$ at the axion minimum $\chi=\pi/2$ for $m=c/2=1/10$.
\end{figure}

\section{Further Examples of Moduli Spaces in $AdS_4$}\label{moreexamples}

In this appendix we give further explicit examples of ${\cal N}=1$
supergravities with degenerate $AdS_4$ backgrounds.
Let us first supply the details of the third example discussed in section~\ref{Examples}, where $K$ and $W$ are given by
\beq
K =  \phi_{1}\bar{\phi}_{1}+\phi_{2}\bar{\phi}_{2}\ ,\qquad
W=c+m \phi_{1}\phi_{2}\ ,
\eeq
with $c$ and $m$ real. One easily computes
\beq\label{eq:Fterms}
D_{\phi_1}W =m\phi_{2}+\bar{\phi}_{1}W\ ,\qquad  D_{\phi_2}W=m\phi_{1}+\bar{\phi}_{2}W
\eeq
and solves $D_{\phi_1}W=D_{\phi_2}W=0$ by parameterizing
$\phi_{1}=r_{1}e^{i(\chi+\rho)},\,\phi_{2}=r_{2}e^{i(\chi-\rho)}$.
The type A trivial solution $\phi_1=\phi_2=0$ is immediately apparent.

If $c$ and $m$ have opposite sign and furthermore $|c| \ge |m|$,
one finds one branch of the type B non-trivial solution,
\be
\label{one} r_1=r_2=\sqrt{-\frac{c}{m}-1}\ ,\quad \chi=0\ , \quad \rho\
\textrm{arbitrary}\ .
\ee
If $c$ and $m$ have the same  sign and again $|c| \ge |m|$,
one finds instead the other branch of the type B non-trivial solution,
\be
\label{two} r_1=r_2 = \sqrt{\frac{c}{m}-1}\ ,\quad \chi=\pi/2\ , \quad \rho\
\textrm{arbitrary}\ .
\ee
These solutions can also be expressed as
$\phi_1=\pm\bar\phi_2$, as in section~\ref{Examples}.
For $|c| = |m|$ these solutions coincide with the trivial solution
$\phi_{1}=\phi_{2}=0$.
We see that $\rho$ is a flat direction for the type B solution with $|c| > |m|$, which can be seen immediately from the fact that
$K$ and $W$  are independent of $\rho$.
Fig.~\ref{Fig.1} shows an example of $V$
%=e^K(K^{i\bar j}D_iW\overline{D_{\bar j}W}-3 |W|^2)$
for $m=c/2$, displaying both solution A at the origin and the second type of the solutions B as saddles at $(\phi_1,\phi_2)=(-1,1)$ and at $(\phi_1,\phi_2)=(1,-1)$.

Let us study the minima of the potential in slightly more detail, as
they
reveal a somewhat unusual structure.
The scalar potential in the variables $r_1, r_2 , \chi$ reads explicitly
\be\begin{aligned}\label{VpotEx2}
V&=e^{r_1^2+r_2^2} \left(c^2 \left(r_1^2+r_2^2-3\right)+2 c m r_1 r_2 \left(r_1^2+r_2^2-1\right) \cos (2\chi)\right.\\
&\qquad\quad\left.+m^2 \left(r_1^4 r_2^2+r_1^2 \left(r_2^4+r_2^2+1\right)+r_2^2\right)\right)\ .
\end{aligned}\end{equation}
The eigenvalues of the matrix of second derivatives at
$\phi_1=\phi_2=0$ are given by
\be
V_{ii}=\left\{0,0,-2 (2 c-m) (c+m),-2 (2 c+m) (c-m)\right\}\ ,
\ee
with a cosmological constant $\vev{V}= -3c^2$.
We see that, as expected, both axions remain flat.

For the non-trivial solutions B the eigenvalues  of the matrix of second derivatives read
 $$V_{B}=\left\{0,-4 m^2 e^{\frac{2 c}{m}-2},4 e^{\frac{2 c}{m}-2}
    (c-m) (2 c+m),4 c m^{-1}  e^{\frac{2 c}{m}-2} (2 c-3 m) (c-m)\right\}$$
 for $c \geq m$ and
 $$V_{B}=\left\{0,-4 m^2 e^{-\frac{2 (c+m)}{m}},-4 e^{-\frac{2
        (c+m)}{m}} (m-2 c) (c+m),-4 c m^{-1}  e^{-\frac{2 (c+m)}{m}} (c+m) (2 c+3 m)\right\}$$
  for $c\leq- m$. The cosmological constant at these extrema turns out to be $\langle V\rangle = -3 m^2 e^{2 \frac{|c| -m}{m}}$. Hence, all of these solutions contain a scalar with negative mass squared, allowed by the Breitenlohner-Freedman (BF)  bound
\be
m_{BF}^2=
-\tfrac{27}{4}e^K|W|^2=-\tfrac{27}{4}e^{\frac{2c}{m}-2}m^2\quad.
\ee
This universal BF scalar is a linear combination of the two radial modes $r_{1,2}$, while the orthogonal linear combination is massive for $|c|>m$. The $\rho$ axion stays massless as expected.

Note that the last eigenvalue gives the mass squared of the $\chi$ axion. We see that for $|c|> \frac 32 m$ this axion is massive,
while its mass vanishes for $|c|=\frac32 m$, and it becomes a BF-stable tachyon for $m<|c|<\frac 32 m$.
This behavior is clear from the scalar potential~\eqref{VpotEx2}.
%The axion $\chi$ only appears through the term $\sim cm (r_1^2+r_2^2-1)\cos(2\chi)$. The points $|c|=\frac 32 m$ constitute the boundary where $r_1^2+r_2^2-1$ flips sign, while the sign of $cm$ is determined by the relative sign of $c$ and $m$. Taken altogether this has the $\chi$-axion mass squared flipping its sign as observed in the eigenvalues of the mass matrices of the type ii) solutions seen above.
%\sda{We've not defined $\Lambda$ up to this point and in any case even in the appendix it is defined only in the decoupling limit. I suggest we take out the first eqn - in any case even if you want it, it needs to be corrected $\Lambda \rightarrow -\Lambda^2$}

Finally, note the unusual vacuum structure of the model. The supersymmetric critical points of type A and B comprise Breitenlohner-Freedman stable tachyonic maxima or saddle points, respectively. We find that for $|c| > m$ the global AdS minima of the model \emph{break} supersymmetry. Hence, the global minima of the scalar potential break supersymmetry, and they have lower vacuum energy than any of the supersymmetric critical points. This is a feature which contradicts intuition from the global case, but is often realized in the context of racetrack models of nonperturbative moduli stabilization in string theory.

%We will continue our list of simple example (from which inductive reasoning led to the general treatment in the main text) with one more example comprising an exact shift symmetry, and two example utilizing a compact $U(1)$ Goldstone symmetry.
%This example borrows the structure of its ingredients from Calabi-Yau compactifications of some string theories, e.g. the heterotic string. The geometric moduli of such compactifications often appear with logarithmic and shift symmetric K\"ahler potentials, while they contribute in the superpotential via exponential dependence in non-perturbative corrections. Guided by this, a very simple toy example reads

As another example  let us consider a supergravity defined by
\begin{equation}
K  =  -\ln(T+\bar{T})\ ,\qquad  W=e^{aT}\ .
\end{equation}
The supersymmetric minimum
$D_{T}W  = %& aW-\frac{1}{T+\bar{T}}W=
W(a-(T+\bar{T})^{-1})=0$ is found for $T+\bar{T}=a^{-1}$ with
$\vev{V}= -3ae$. It only exists if $a$ is real and positive,
and then $\Im T$ is the modulus of this $AdS_4$ background.
This can be generalized for a generic  $K=K(T+\bar{T})$ with a
shift symmetry, as long as
$K_T\ne 0,K_{T\bar T}>0$. In this case  $\Re T$ is fixed by
$K_T=-a$.
In string theory %$T+\bar T$ should be positive (being the
%                           %size of a cycle, or the inverse string
%                           %coupling) but this would result in
$a>0$  %so the expression for $W$  would not be a valid pure
%       %non-perturbative term per se.
does not easily appear, as $W$ then diverges in the large $T$ limit.
However, a superpotential $W=A\exp (-b(S-aT/b))$ with $b,a>0$ can arise,
for example, in heterotic backgrounds, where the second term in the
exponent can be a threshold correction and $S$ is the dilaton.

As a further example consider
\begin{equation}
K  =  \phi\bar{\phi}\ ,\qquad W=a\phi^{b}\ ,
\end{equation}
which has an R-symmetry.
The supersymmetric minimum
$D_{\phi}W
=a\phi^{b-1}(b+\bar{\phi}\phi) =0$
is $AdS_4$ for $b<0, a\neq0$ and $\phi\bar{\phi}=|b|$. In this case
the phase of $\phi$ is the modulus and
we have $\vev{V}=-3e^{|b|}|a|^{2}|b|^{b}$.
Note however that the superpotential needs to be singular at the origin.
% in order to make the example work.

Our final example realizes  a compact $U(1)$ moduli space  but is a bit more involved. This is due to the fact that this case is set up to avoid singularities in field space as well as the use of an arbitrary constant term in the superpotential. Consider
\be
K=\phi\bar{\phi}+\chi\bar{\chi}+X\bar{X}\ , \qquad
W=X(\chi\phi-\mu^{\tilde{}2})+m\phi\chi\ ,
\ee
which are invariant under $\phi\to e^{i\alpha}$, $\chi\to e^{-i\alpha}\chi$.
One easily computes
\be\begin{aligned}
D_XW&=\chi\phi-\mu^2+\bar X W\ ,\\
D_\phi W&=(m+X)\chi+\bar\phi W\ ,\\
D_\chi W&=(m+X)\phi+\bar\chi W\ .
\end{aligned}\ee
This theory has a supersymmetric $AdS_4$ background with
%a dynamically generated cosmological constant  and
a flat direction corresponding
to the $U(1)$ symmetry.
The full solution
can be obtained analytically but is not
particularly illuminating.
Instead we display the solution  for $m\ll \mu$, which captures the
essential features:
\be\begin{aligned}
\langle |\chi|\rangle=\langle
|\phi|\rangle&=\mu\cdot\left[1+\frac{m^2}{2}\;(1+\mu^2)\right]+{\cal
  O}(m^3)\ ,\\
\langle X\rangle&=-(1+\mu^2)\,m+{\cal O}(m^3)\ ,\\
\langle W\rangle&=m\mu^2\cdot\left[1-m^2\mu^2(1+\mu^2)\right]+{\cal
  O}(m^4)\ ,
\end{aligned}\ee
where a common phase of $\phi$ and $\chi$ is left undetermined.

\bibliography{GLSTV}
\providecommand{\href}[2]{#2}\begingroup\raggedright\endgroup

\end{document}